\documentclass[journal]{IEEEtran}

\usepackage[margin=1in]{geometry}         % 1-inch margins
\usepackage{setspace}                     % Line spacing
\usepackage[all]{nowidow}                 % Prevents single lines at page start/end
\usepackage{microtype}                    % Better character protrusion & expansion
\usepackage{indentfirst}                  % Indent the first paragraph after sections
\usepackage{footmisc}                     % Footnote control
\usepackage[T1]{fontenc}                  % Output font encoding
\usepackage[utf8]{inputenc}               % Input character encoding
\usepackage[english]{babel}               % Language-specific rules
\usepackage{csquotes}                     % Context-sensitive quotes
\usepackage{comment}

\usepackage{amsmath,mathtools}            % Enhanced math environments
\usepackage{bm}                           % Proper bold math (retains italics)
\usepackage{bbm}                          % Blackboard bold for digits
\usepackage{graphicx}                     % Support for images
\usepackage{subcaption}                   % Subfigures (1a, 1b, etc.)
\usepackage{float}                        % Better control over float placement [H]
\usepackage{booktabs}                     % Publication quality tables
\usepackage{threeparttable}               % Tables with integrated notes
\usepackage{tabularx}                     % Flexible-width columns
\usepackage{longtable}                    % Multi-page tables
\usepackage{tabularray}                   % Modern table styling
  \UseTblrLibrary{booktabs}               % Allows booktabs commands in tabularray
\usepackage{adjustbox}                    % Resizing/rotating
\usepackage{pdflscape}                    % Landscape pages
\usepackage{rotating}                     % Sideways elements
\usepackage{enumitem}
\setlist{nosep}                           % Tighter spacing for lists
\usepackage{multirow}

\usepackage{siunitx}                      % Aligning numbers by decimal point
\usepackage{cite}

\usepackage{fancyhdr}
\usepackage[hidelinks]{hyperref}          % Clickable links
\usepackage[nameinlink,noabbrev]{cleveref} % Smart referencing (Table 1, Fig 2)

\title{\textbf{Analyzing the Impact of Release Season and Production Budget on Movie Revenue and Profitability}}

\author{
\IEEEauthorblockN{
Mohammad Jalili Torkamani\IEEEauthorrefmark{1},
Pedro Gomes\IEEEauthorrefmark{2},
Amirmohammad Sadeghnejad\IEEEauthorrefmark{3},
Jason Le\IEEEauthorrefmark{1}
}

\IEEEauthorblockA{\IEEEauthorrefmark{1}
School of Computing (Department of Engineering)\\
University of Nebraska--Lincoln\\
Lincoln, NE, USA\\
\{mjalilitorkamani2, jle19\}@huskers.unl.edu
}

\IEEEauthorblockA{\IEEEauthorrefmark{2}
Department of Supply Chain Management and Analytics\\
University of Nebraska--Lincoln\\
Lincoln, NE, USA\\
pgomes2@huskers.unl.edu
}

\IEEEauthorblockA{\IEEEauthorrefmark{3}
Department of Civil and Environmental Engineering\\
University of Nebraska--Lincoln\\
Lincoln, NE, USA\\
asadeghnejad2@huskers.unl.edu
}
}

\begin{document}

\maketitle

\section{Abstract}
The film industry is characterized by significant financial uncertainty, where large production investments do not always guarantee commercial success. This study analyzes the relationship between release season, production budget, and movie financial performance using the Full TMDB Movies Dataset 2024. A data mining framework incorporating association rule mining, clustering, machine learning, and SHAP analysis was applied to identify key drivers of revenue and profitability.

The results show that release season has limited predictive influence on revenue and return on investment (ROI). In contrast, production budget, popularity, and audience ratings are significantly more influential. Association rule mining revealed that high-budget films with poor ratings are strongly associated with negative ROI outcomes. Random Forest regression achieved substantially stronger predictive performance than Decision Tree regression, with an $R^2$ value of 0.652. SHAP analysis further confirmed that budget and popularity are the dominant predictors of box office revenue, while timing-related variables contribute minimally.

These findings suggest that financial success in the film industry is driven more by production investment and market attention than by seasonal release strategies, providing practical insights for budgeting, release planning, and financial risk management.

\section{Introduction}

The film industry is characterized by substantial financial uncertainty. Producing a movie requires large upfront investments in development, production, and marketing, while financial outcomes, typically measured as box office revenue, remain difficult to predict. Studios must determine how much to invest and when to release films with limited empirical evidence on how key factors, such as release timing and budget, influence performance. Common industry beliefs suggest that winter releases benefit from holiday demand and awards season exposure, and that higher budgets lead to greater revenue. However, these assumptions may not hold consistently across different market conditions.

From a data mining and business analytics perspective, the central challenge is to determine whether systematic relationships exist between release season, production budget, and performance measures such as revenue, ratings, and profitability. In the absence of rigorous analysis, studios and investors face elevated decision risk, potentially leading to inefficient resource allocation and suboptimal release strategies.

Several factors complicate this problem. Films released within the same season compete for audience attention, making it difficult to isolate true seasonal effects. Larger production budgets increase financial exposure, as underperforming films can result in substantial losses. At the same time, revenue and ratings are shaped by multiple interacting variables, limiting the predictive power of any single factor. To address these challenges, this study applies data preprocessing, statistical analysis, and predictive modeling techniques to systematically examine these relationships.

This study contributes practical insights for decision-making in the film industry. By analyzing how release season and production budget relate to revenue, ratings, and profitability, the results provide evidence to support more informed strategic decisions. Identifying whether certain seasons consistently yield stronger outcomes can help studios optimize release timing and manage competitive pressures. Evaluating the performance of low- and high-budget films can improve resource allocation decisions. In addition, developing predictive models for revenue offers a data-driven framework for estimating financial outcomes prior to release. These insights support more effective planning in scheduling, budgeting, and risk management within the film market.

\section{Theoretical Background}

Prior research has examined a range of factors influencing film success. \cite{einav2007seasonality} analyzed the role of release season and its effect on movie performance, finding that seasonal peaks are not driven by higher consumer demand but rather by strategic behavior from distributors. In particular, higher-quality films tend to be released during peak periods, such as the winter season, creating a self-reinforcing cycle in which stronger films perform better because they are concentrated in highly competitive release windows.

Using a data-driven approach, \cite{xiao2024modeling} applied logistic regression, support vector machines, and random forest models to predict box office success. Their results indicate that production budget is a significant positive predictor of performance, but with diminishing returns beyond a certain threshold. This suggests that higher spending increases the likelihood of success, but does not guarantee profitability.

Taken together, these studies highlight that film performance is influenced by multiple interacting factors rather than any single attribute, underscoring the complexity of predicting financial outcomes in the film industry.

The economic analysis of the motion picture industry reveals a complex ecosystem in which box office revenue and profitability represent distinct concepts \cite{lash2015}. While public attention often focuses on gross box office receipts, financial performance is more accurately captured through measures such as Return on Investment (ROI) and net profits \cite{lash2015}. In this context, high box office revenue does not necessarily imply strong financial returns, as large production and marketing costs can substantially reduce overall profitability \cite{lash2015}.

Traditional models of movie success have frequently emphasized the role of actor "star power" as a primary determinant of performance. However, empirical evidence suggests that this effect is weaker than commonly assumed. Instead, variables such as the historical profitability of actor-director collaborations and genre-specific market trends provide stronger predictive power for ROI \cite{lash2015}. From a production perspective, teams that combine domain-specific expertise with diverse creative input tend to achieve higher financial performance, highlighting the importance of collaborative dynamics in shaping outcomes \cite{lash2015}.

The evolution of distribution channels has also reshaped the structure of movie revenue. The "long tail" theory posits that digital platforms would lead to increased consumption of niche content \cite{drost2021}. However, evidence from the film industry suggests the opposite effect: the rise of streaming platforms has reinforced revenue concentration among a small number of blockbuster films \cite{drost2021}. This pattern is consistent with the "superstar theory," which argues that reduced distribution constraints amplify consumer focus on high-profile content \cite{drost2021}.

In the modern marketing environment, digital engagement has become a key driver of audience behavior. For example, trailer viewership on platforms such as YouTube has been shown to be a strong predictor of box office performance, reflecting the role of pre-release attention in shaping demand \cite{seay2025movie}. At the same time, the measurement of profitability is complicated by industry practices commonly referred to as "Hollywood accounting," whereby studios use contractual and accounting mechanisms to redefine net profits and delay reported break-even points \cite{daniels1998movie}. As a result, even films with substantial box office revenue may not report positive net profits under contractual definitions \cite{daniels1998movie}.

\section{Objectives}

This project is organized into four main objectives, consisting of an initial data preprocessing stage followed by three data mining tasks, as outlined below.

\paragraph{Objective 1: Data Preprocessing}
Prepare the dataset to ensure accuracy and consistency in subsequent analyses. This includes handling missing values, correcting inconsistencies, defining release seasons, constructing profitability measures (profit and ROI), and transforming variables into appropriate formats for statistical analysis and modeling.

\paragraph{Objective 2: Association Pattern Mining}
Examine whether release season is associated with revenue and audience ratings by identifying potential patterns and differences across seasonal categories. This objective incorporates exploratory data analysis and association rule mining (e.g., Apriori) to evaluate the extent to which release timing influences key performance indicators.

\paragraph{Objective 3: Season-Profitability Knowledge Discovery}
Identify underlying patterns in film profitability and assess whether these patterns vary across seasons. This involves applying segmentation and classification techniques (e.g., clustering and Random Forest) to evaluate whether financial performance profiles differ systematically by release period.

\paragraph{Objective 4: Uncovering Movie Revenue Key Drivers}
Analyze the key determinants of film revenue and evaluate whether low-budget films can achieve high financial performance. This objective includes the development of predictive models (e.g., Decision Tree and Random Forest regression) and the use of SHAP analysis to quantify the relative importance of features such as budget, popularity, and ratings.

\section{Datasets}

This project uses the Full TMDB Movies Dataset 2024 (1M Movies) compiled by Asaniczka. The dataset is based on information collected from The Movie Database (TMDB) and includes movie records spanning from the early 1900s to 2024. It is publicly available and updated regularly. The dataset contains 1,364,577 movie records and 24 attributes per movie. It covers more than a century of film releases across multiple countries and languages. The large size of the dataset allows for comprehensive seasonal, financial, and performance-based analysis.

The dataset includes a combination of:

\begin{itemize}
    \item Date variables (e.g., release\_date)
    \item Categorical/character variables (e.g., title, genres, production\_companies)
    \item Logical variables (e.g., adult)
    \item Numerical variables (e.g., revenue, budget, runtime, vote\_average, vote\_count, popularity)
\end{itemize}

The primary variables used in this project include:
\begin{itemize}
    \item release\_date (used to define seasons)
    \item revenue (box office earnings)
    \item budget (production cost)
    \item vote\_average and vote\_count (audience evaluation metrics)
    \item popularity (overall interest level)
\end{itemize}

\section{Data Preprocessing}

To prepare the dataset for analysis, a structured preprocessing procedure was implemented in R using the TMDB dataset. An initial exploratory assessment was conducted using the skim function to examine variable types, missing values, and overall data structure. This step helped identify attributes with substantial missing data as well as variables not directly relevant to the research objectives. Table~\ref{tab:tmdb_raw} summarizes the attributes, their data types, and completeness rates.

\begin{table}[htbp]
\setstretch{1}
\caption{TMDB Dataset Attributes and Completeness}
\label{tab:tmdb_raw}
\centering
\footnotesize
\resizebox{\columnwidth}{!}{
\begin{tabular}{llrr}
\hline
\textbf{Variable} & \textbf{Type} & \textbf{Missing} & \textbf{Completeness} \\
\hline
release\_date & Date & 288,187 & 78.88\% \\
title & String & 22 & 100.00\% \\
status & String & 0 & 100.00\% \\
backdrop\_path & String & 1,020,721 & 25.20\% \\
homepage & String & 1,223,634 & 10.33\% \\
imdb\_id & String & 708,461 & 48.08\% \\
original\_language & String & 0 & 100.00\% \\
original\_title & String & 18 & 100.00\% \\
overview & String & 309,880 & 77.29\% \\
poster\_path & String & 473,872 & 65.27\% \\
tagline & String & 1,174,547 & 13.93\% \\
genres & String & 590,805 & 56.70\% \\
production\_companies & String & 777,271 & 43.04\% \\
production\_countries & String & 650,697 & 52.32\% \\
spoken\_languages & String & 625,381 & 54.17\% \\
keywords & String & 1,022,063 & 25.10\% \\
adult & Logical & 0 & 100.00\% \\
id & Numeric & 0 & 100.00\% \\
vote\_average & Numeric & 0 & 100.00\% \\
vote\_count & Numeric & 0 & 100.00\% \\
revenue & Numeric & 0 & 100.00\% \\
runtime & Numeric & 0 & 100.00\% \\
budget & Numeric & 0 & 100.00\% \\
popularity & Numeric & 0 & 100.00\% \\
\hline
\end{tabular}
}
\end{table}

Although the dataset is well maintained, several descriptive attributes contain substantial missing values, including "homepage", "tagline", "backdrop\_path", and "keywords". In contrast, key financial and performance variables such as revenue, budget, ratings, and popularity exhibit full completeness.

Following this assessment, a structured data transformation pipeline was applied. First, duplicate movie entries were addressed by grouping observations by "imdb\_id" and retaining the record with the highest available revenue and budget values. This ensures that each film is uniquely represented while preserving the most complete financial information.

Next, the dataset was filtered to remove non-relevant observations. Adult films were excluded, and only films released after May 16, 1929 (the date of the first Academy Awards) were retained to ensure temporal consistency. Several non-essential and metadata variables, such as identifiers, textual descriptions, and image-related attributes, were removed to reduce dimensionality and focus on variables relevant to financial performance.

The remaining variables were then standardized and renamed for clarity, including financial attributes (budget, revenue), temporal features (season, days to Oscar, days to holiday), and performance indicators (rating, popularity). Derived variables were also constructed to capture financial outcomes. In particular, profit was defined as the difference between revenue and budget, and profit ROI was computed as a normalized measure of financial performance relative to investment.

To support different analytical tasks, two datasets were created. A broad dataset (movies\_db\_broad) retained all variables, including financial measures, and was restricted to complete observations using list-wise deletion. A narrow dataset (movies\_db\_narrow) excluded financial variables such as revenue, budget, and ROI, and was similarly restricted to complete cases. These datasets were used for subsequent unsupervised and supervised learning tasks.

Finally, an additional simplified dataset was constructed for baseline analyses. In this version, non-essential variables were removed, key variables were renamed, and observations with missing values in primary attributes were excluded. These preprocessing steps ensure that the dataset is clean, consistent, and suitable for statistical analysis and predictive modeling.

Although several descriptive variables contain missing values, key financial and rating variables such as revenue, budget, vote\_average, vote\_count, and popularity exhibit full completeness in the raw dataset. Nevertheless, the presence of missing values in other fields necessitated preprocessing prior to analysis to ensure data consistency.

Second, non-essential variables were removed to improve analytical clarity. Specifically, backdrop\_path, homepage, imdb\_id, and poster\_path were excluded, as these fields primarily contain metadata or external identifiers that are not directly related to financial performance or release timing.

Next, a new variable, profit, was created to measure financial performance. Profit was defined as the difference between box office revenue and production budget:
\begin{equation}
    Profit = Revenue - Budget
    \label{eq:profit}
\end{equation}

Based on profit, a normalized measure of financial performance, Profit ROI, was also constructed to capture return on investment relative to the initial production budget:
\begin{equation}
    Profit\ ROI = \frac{Revenue - Budget}{Budget}
    \label{eq:profitROI}
\end{equation}

Profit ROI was preferred over raw revenue because it enables comparison across films with substantially different budget scales. Using revenue alone would bias the analysis toward high-budget productions, potentially masking the relative financial success of lower-budget films. By normalizing financial outcomes relative to investment, Profit ROI provides a more consistent measure of profitability across movies with varying production costs.

In later stages of the analysis, Profit ROI was used as the primary financial performance metric for profitability evaluation, clustering, and association pattern mining.

Finally, missing values for the primary variables of interest were handled using list-wise deletion (drop\_na()), retaining only complete observations for the selected financial, rating, and release date attributes. This approach ensured consistency in statistical testing and predictive modeling by eliminating incomplete records that could bias the analysis.

The resulting cleaned dataset was used for seasonal comparison analysis, profitability evaluation, and predictive modeling tasks. Table~\ref{tab:tmdb_cleaned} summarizes the updated attribute completeness after preprocessing.

\begin{table}[htbp]
\setstretch{1}
\caption{TMDB Cleaned Dataset: Variable Types and Completeness}
\label{tab:tmdb_cleaned}
\centering
\footnotesize
\resizebox{\columnwidth}{!}{
\begin{tabular}{llrr}
\hline
\textbf{Variable} & \textbf{Type} & \textbf{Missing} & \textbf{Completeness} \\
\hline
released & Date & 0 & 100.00\% \\
title & String & 5 & 100.00\% \\
status & String & 0 & 100.00\% \\
original\_language & String & 0 & 100.00\% \\
original\_title & String & 4 & 100.00\% \\
overview & String & 194,357 & 81.94\% \\
tagline & String & 920,488 & 14.48\% \\
genres & String & 361,994 & 66.37\% \\
production\_companies & String & 522,245 & 51.48\% \\
production\_countries & String & 400,573 & 62.79\% \\
spoken\_languages & String & 395,953 & 63.21\% \\
keywords & String & 757,755 & 29.60\% \\
adult & Logical & 0 & 100.00\% \\
id & Numeric & 0 & 100.00\% \\
rating & Numeric & 0 & 100.00\% \\
vote\_count & Numeric & 0 & 100.00\% \\
revenue & Numeric & 0 & 100.00\% \\
runtime & Numeric & 0 & 100.00\% \\
budget & Numeric & 0 & 100.00\% \\
popularity & Numeric & 0 & 100.00\% \\
profit & Numeric & 0 & 100.00\% \\
profit\_ROI & Numeric & 0 & 100.00\% \\
\hline
\end{tabular}
}
\end{table}

\subsection{Descriptive Statistics}

\begin{figure}[H]
\centering
\includegraphics[width=\columnwidth]{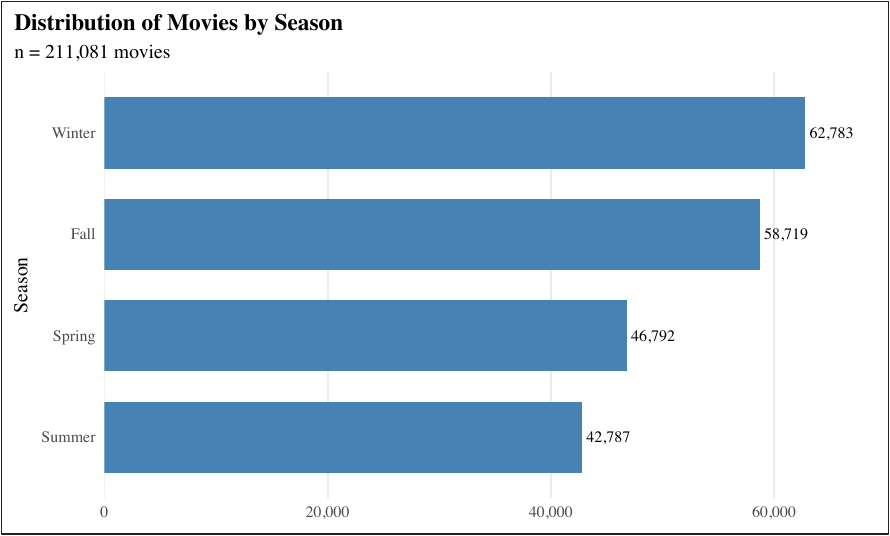}
\caption{Distribution of Movies by Season}
\label{fig_season_dist}
\end{figure}

Figure~\ref{fig_season_dist} presents the distribution of films across the four release seasons. The dataset is relatively well balanced, with all seasons containing a substantial number of observations. Spring and Fall exhibit the highest counts, followed by Summer, while Winter contains a slightly smaller but still comparable number of films.

This distribution indicates that no single season dominates the dataset, reducing concerns about class imbalance in subsequent analyses. The relatively even representation across seasons is particularly important for both unsupervised and supervised learning tasks, as it ensures that observed patterns are not driven by disproportionate sample sizes.

At the same time, the absence of strong imbalance suggests that any differences in financial performance across seasons are unlikely to be attributed to data availability alone. Instead, this balanced structure provides a suitable foundation for evaluating whether meaningful seasonal effects exist in revenue, ratings, and profitability.

\begin{figure}[H]
\centering
\includegraphics[width=\columnwidth]{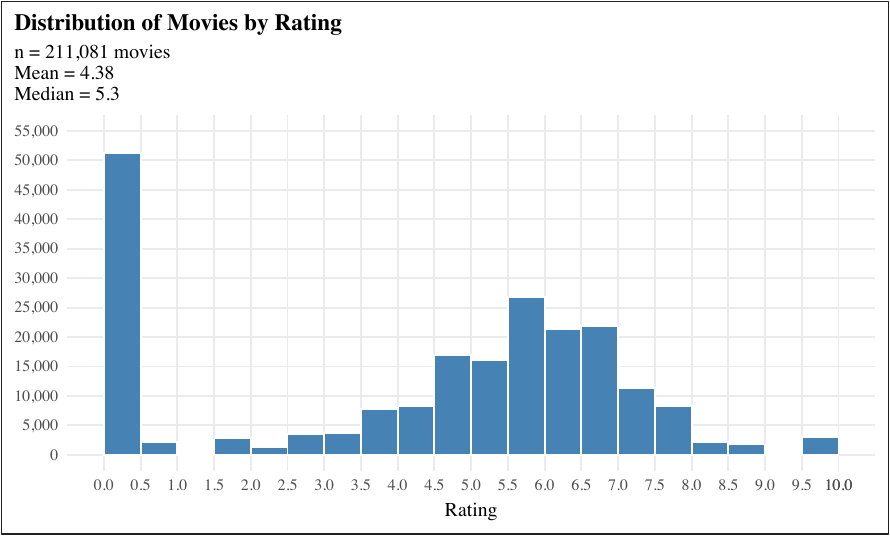}
\caption{Distribution of Movie Ratings}
\label{fig_rating_dist}
\end{figure}

Figure~\ref{fig_rating_dist} presents the distribution of movie ratings across the dataset. The distribution is concentrated within a moderate range, with most films receiving ratings between approximately 4 and 7. The mean rating is 4.38, while the median is 5.3, indicating a slight left skew driven by a larger number of lower-rated films.

This pattern suggests that extremely high-rated films are relatively uncommon, and that the majority of movies receive average to moderately positive evaluations. The limited dispersion in ratings indicates that audience and critic evaluations are relatively compressed compared to financial outcomes, which tend to exhibit much greater variability.

From an analytical perspective, this distribution implies that rating alone may have limited power in explaining large differences in revenue. While higher ratings are generally associated with improved performance, the relatively narrow range of values suggests that other factors, such as budget and popularity, are likely to play a more dominant role in driving financial outcomes.

\begin{figure}[H]
\centering
\includegraphics[width=\columnwidth]{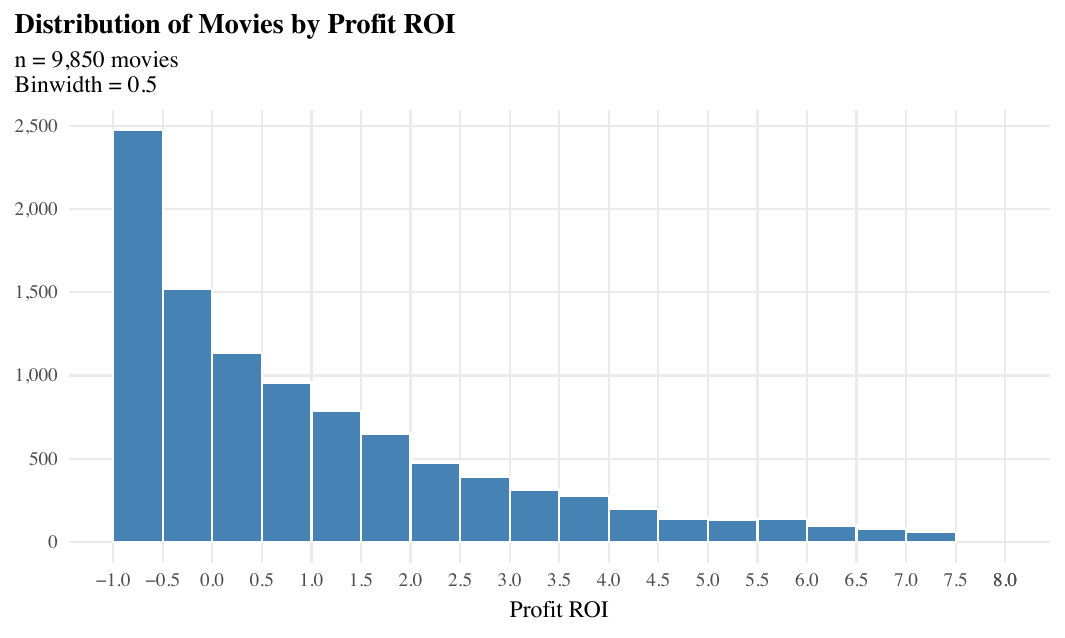}
\caption{Distribution of Profit ROI after Preprocessing}
\label{fig_roi_dist}
\end{figure}

Figure~\ref{fig_roi_dist} presents the distribution of Profit ROI after preprocessing. The distribution remains highly skewed, with a large concentration of films clustered around low or near-zero returns, and a long right tail representing a smaller number of highly profitable movies.

A substantial portion of films exhibit negative or marginal ROI, indicating that many productions fail to recover their initial investment. At the same time, the presence of extreme positive values highlights the existence of a limited number of highly successful films that generate disproportionately large returns. This asymmetry reflects the well-documented “hit-driven” nature of the film industry, where a small number of blockbuster successes account for a large share of total revenue.

Despite the removal of extreme outliers during preprocessing, the distribution retains significant variability, suggesting that financial outcomes remain inherently uncertain and difficult to predict. From a modeling perspective, this wide dispersion implies that simple linear relationships may be insufficient to capture revenue dynamics, motivating the use of more flexible methods such as tree-based models and ensemble techniques.

\section{Association Pattern Mining}

This objective aims to identify hidden relationships in the dataset using association rule mining. The Apriori algorithm was applied to uncover patterns across financial, temporal, and structural attributes.

\subsection{Outlier Removal}

\begin{figure}[!t]
\centering
\includegraphics[width=\columnwidth]{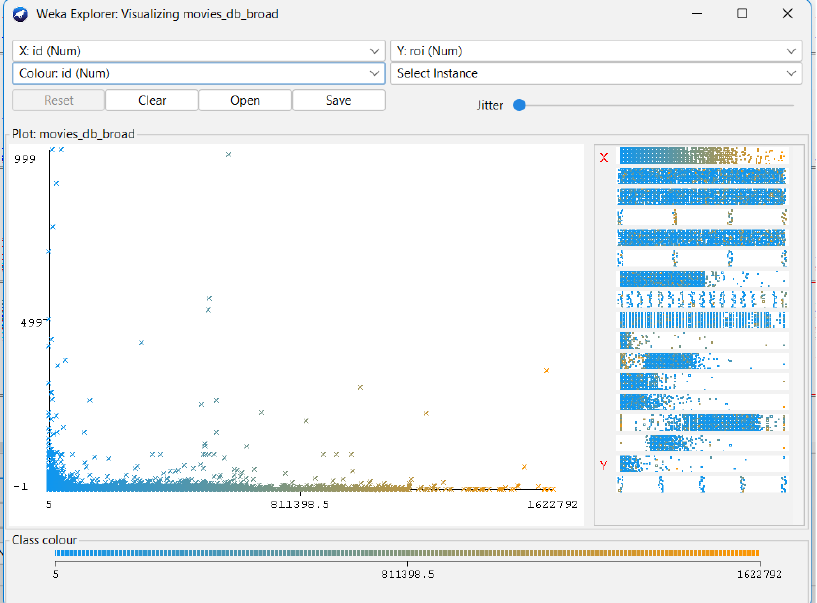}
\caption{Return on Investment Distribution after Outlier Removal}
\label{fig_plot_roi}
\end{figure}

Initial exploration revealed substantial skewness in both budget and ROI distributions. Extremely low reported budgets (e.g., values near 1) generated implausibly large ROI values, while certain films exhibited extreme profitability. To address this, observations with budgets below \$1,000 were removed to avoid distortion in the analysis. In addition, extreme outliers such as \textit{The Blair Witch Project}, which achieved exceptionally high ROI, were excluded to stabilize the distribution and improve interpretability. The resulting ROI distribution is shown in Figure~\ref{fig_plot_roi}.

A minor preprocessing step was also performed to standardize the holiday variable by removing redundant labels such as “(Observed)”, which otherwise created duplicate categories.

\subsection{Attribute Construction}

Because Apriori requires categorical inputs, continuous variables were discretized into bins to enable rule generation while preserving interpretability:

\begin{itemize}
    \item ROI: Negative (<0), Low (0--1), Medium (1--3), High (3--10), Very High (>10)
    \item Rating: Low, Mid-Low, Mid-High, High (equal-frequency discretization)
    \item Days to Holiday: Close ($\leq$7), Mid (8--30), Far (>30)
    \item Runtime: Short ($\leq$97), Average (97--113), Long (>113)
    \item Budget: Low, Medium, High, Very High (equal-frequency discretization)
    \item Season Period: Early, Mid, Late within each season
\end{itemize}

\subsection{Initial Apriori Results}

The algorithm was first executed with a minimum support of 0.05 and minimum confidence of 0.7. Although the resulting rules exhibited high confidence, they were not substantively meaningful. Most rules reflected deterministic relationships between calendar-based variables, such as:

\begin{itemize}
    \item Spring\_Early $\rightarrow$ Washington's Birthday
    \item Winter\_Early $\rightarrow$ Christmas Day
\end{itemize}

These associations were mechanically induced by overlapping definitions of time-related attributes rather than reflecting underlying behavioral or financial patterns in the data.

\subsection{Model Refinement and Parameter Tuning}

To reduce redundancy, the season\_period attribute was removed from the analysis. However, rerunning the Apriori algorithm with the same thresholds produced no rules, indicating that the original criteria were overly restrictive once redundant attributes were excluded.

The thresholds were then relaxed to a minimum support of 0.03 and minimum confidence of 0.6. Under these updated settings, the algorithm identified several meaningful associations, which can be grouped into three categories.

\noindent \textbf{Category 1: Structural Relationships}

\begin{itemize}
    \item High rating \& very high budget $\rightarrow$ long runtime (confidence = 0.69, lift = 2.05)
    \item Low rating \& low budget $\rightarrow$ short runtime (confidence = 0.61, lift = 1.98)
    \item High rating \& high budget $\rightarrow$ long runtime (confidence = 0.60, lift = 1.78)
\end{itemize}

\begin{figure}[!t]
\centering
\includegraphics[width=\columnwidth]{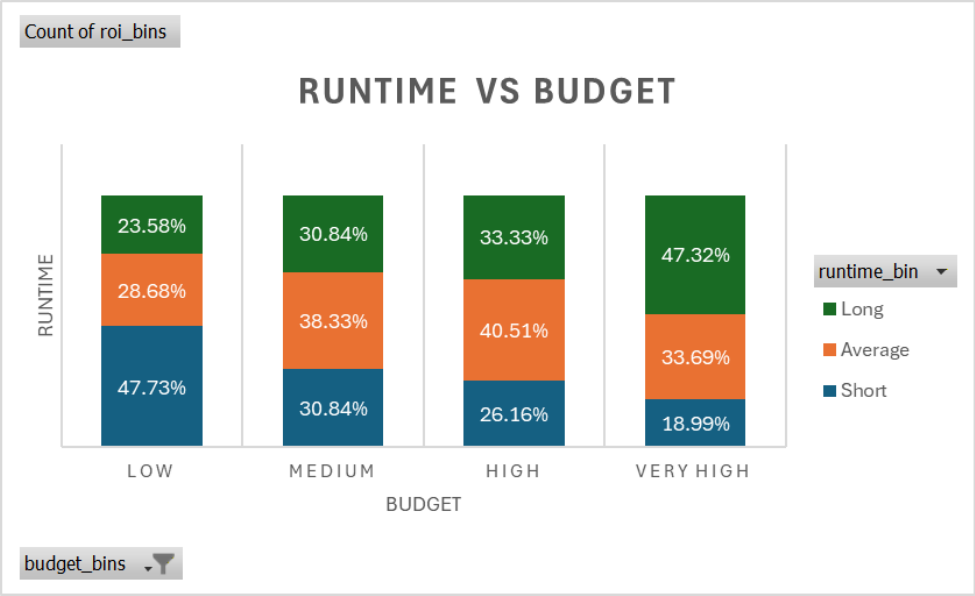}
\caption{Runtime Distribution across Budget Bins}
\label{fig_plot_runtime_budget}
\end{figure}

These results indicate that higher-budget and better-rated films tend to have longer runtimes, while lower-budget films are more likely to be shorter. This pattern is consistent with production constraints and resource availability, where larger productions have greater capacity to support extended content. Figure~\ref{fig_plot_runtime_budget} supports this relationship by showing an increasing share of long films as budget rises.

\noindent \textbf{Category 2: Temporal Relationships}

Temporal rules primarily captured redundant relationships among date-derived variables (e.g., proximity to holidays). These patterns do not provide substantive insights, suggesting that release timing has limited direct association with ROI or ratings in this dataset.

\noindent \textbf{Category 3: Financial Relationships}

\begin{itemize}
    \item Low rating \& high budget $\rightarrow$ negative ROI (confidence = 0.62, lift = 1.76)
\end{itemize}

\begin{figure}[!t]
\centering
\includegraphics[width=\columnwidth]{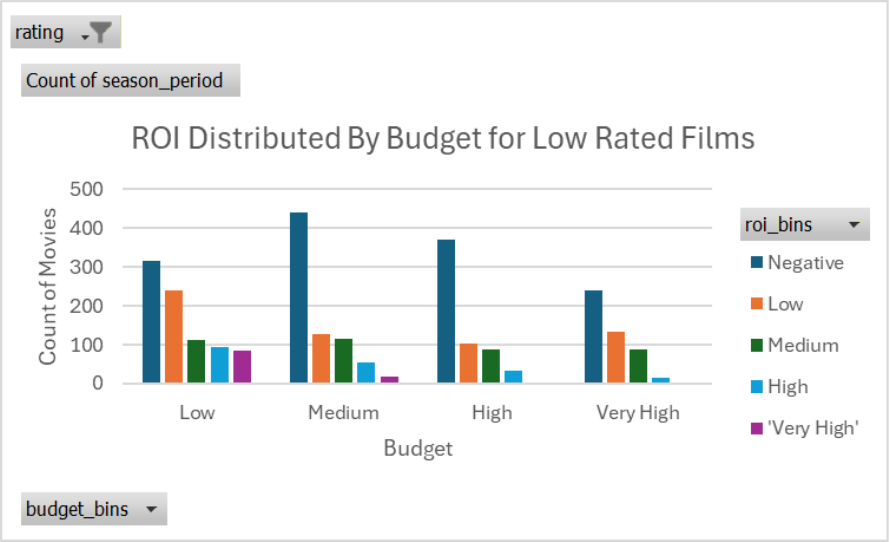}
\caption{ROI Distribution across Budget Bins for Low Rated Films}
\label{plot_roi_budget_low}
\end{figure}

\begin{figure}[!t]
\centering
\includegraphics[width=\columnwidth]{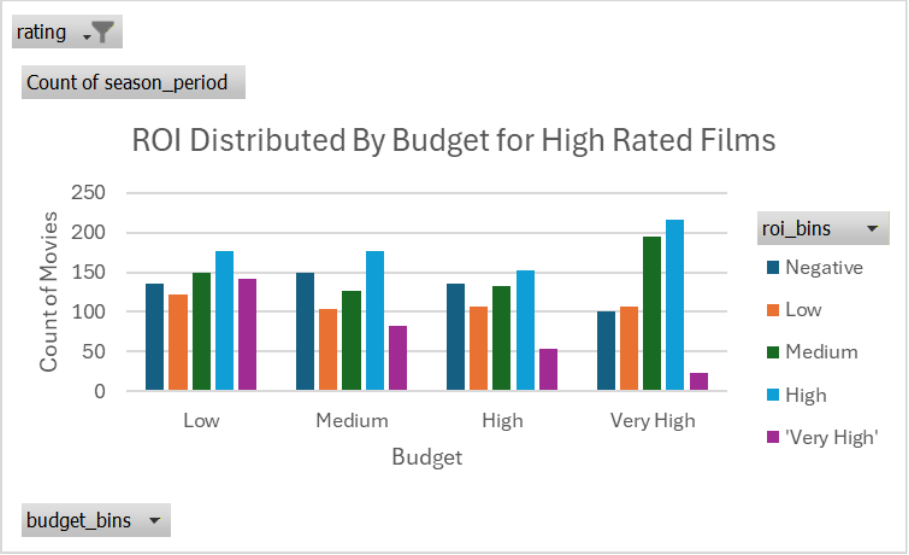}
\caption{ROI Distribution across Budget Bins for High Rated Films}
\label{plot_roi_budget_high}
\end{figure}

This is the most substantively meaningful rule identified. It indicates that high-budget films with poor audience reception are more likely to generate financial losses. Figures~\ref{plot_roi_budget_low} and \ref{plot_roi_budget_high} are consistent with this finding, showing that low-rated films exhibit a substantially higher proportion of negative ROI outcomes compared to higher-rated films.

Figure~\ref{plot_roi_budget_high} further highlights the complexity of predicting movie ROI, as a wide range of outcomes is observed across all budget categories.

\subsection{Profit ROI Prediction Focus}

To further refine the analysis, class association rule mining was implemented by setting ROI as the target variable. Under the same thresholds, only one rule satisfied the criteria:

\begin{itemize}
    \item Low rating \& high budget $\rightarrow$ negative ROI (confidence = 0.62, lift = 1.76)
\end{itemize}

The absence of additional rules suggests that the available features provide limited predictive power for financial outcomes. The identified association patterns are relatively sparse, with structural relationships largely reflecting expected industry characteristics and temporal variables contributing minimal explanatory value. The most robust finding is the negative association between poor audience reception and financial performance for high-budget films, indicating that large investments amplify downside risk when films are poorly received.

At the same time, the limited number of meaningful rules underscores the inherent complexity of predicting movie success, as financial outcomes depend on multiple interacting factors that are not fully captured by the available variables. These results indicate that while association rule mining can reveal interpretable relationships, its ability to explain or predict financial performance in this context remains limited.

\section{Season-Profitability Knowledge Discovery}

Our analysis utilized the movies\_db\_broad dataset, which initially contained 10,895 records. Records with missing values and irrelevant attributes were removed prior to this stage to ensure a complete and consistent dataset. The dimensionality of the dataset was refined to an attribute space including season, budget, and revenue.

To improve model performance and robustness, noise reduction was implemented through the removal of extreme outliers using an Interquartile Range (IQR) unsupervised filter in Weka. This step mitigates the influence of extreme observations on the analysis. Specifically:

\begin{itemize}
    \item Attribute indices for profit and ROI were targeted for outlier mitigation.
    \item Extreme values were flagged and removed from the instance set.
    \item Auxiliary attributes generated during the filtering process were subsequently discarded.
\end{itemize}

After preprocessing, the dataset was reduced to 8,723 records for use in both unsupervised and supervised learning tasks. Summary statistics for the key financial attributes are presented in Table~\ref{tab:financial_stats}.

\begin{table}[H]
\centering
\caption{Financial Attribute Statistics}
\label{tab:financial_stats}
\begin{tabular}{llr}
\toprule
\textbf{Attribute} & \textbf{Statistic} & \textbf{Value} \\
\midrule
\multirow{4}{*}{Profit} 
& Minimum & -98,491,868 \\
& Maximum & 127,469,017 \\
& Mean    & 8,171,419.691 \\
& StdDev  & 27,801,401.268 \\
\midrule
\multirow{4}{*}{ROI} 
& Minimum & -1 \\
& Maximum & 4.994 \\
& Mean    & 0.558 \\
& StdDev  & 1.405 \\
\bottomrule
\end{tabular}
\end{table}

\begin{figure}[!t]
\centering
\includegraphics[width=\columnwidth]{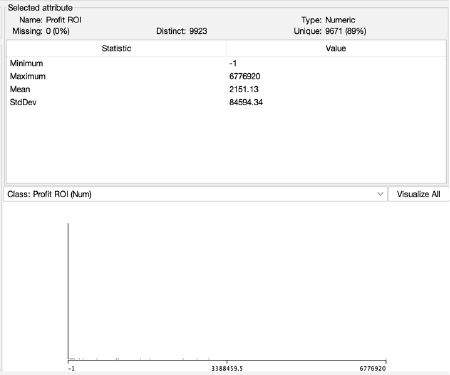}
\caption{ROI Distribution Before Outlier Removal}
\label{fig:roi_outliers}
\end{figure}

\begin{figure}[!t]
\centering
\includegraphics[width=\columnwidth]{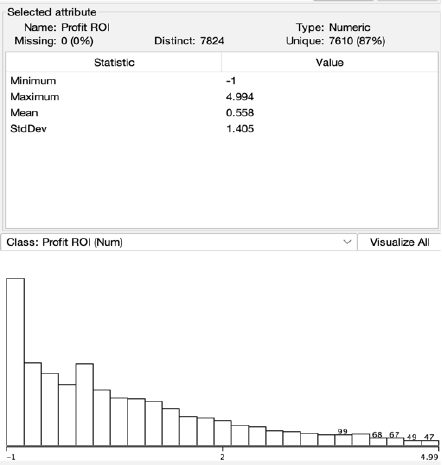}
\caption{ROI Distribution After Outlier Removal}
\label{fig:roi_wo_outliers}
\end{figure}

\begin{figure}[!t]
\centering
\includegraphics[width=\columnwidth]{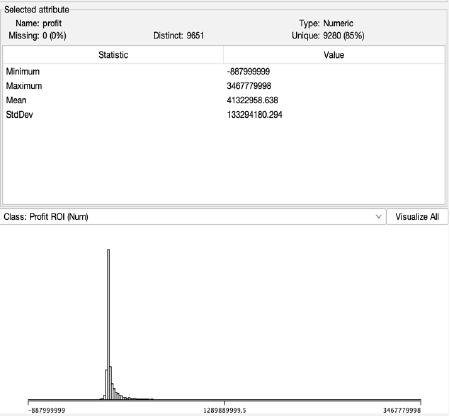}
\caption{Profit Distribution Before Outlier Removal}
\label{fig:profit_outliers}
\end{figure}

\begin{figure}[!t]
\centering
\includegraphics[width=\columnwidth]{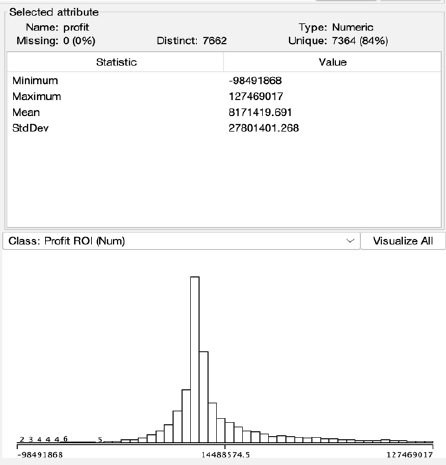}
\caption{Profit Distribution After Outlier Removal}
\label{fig:profit_wo_outliers}
\end{figure}

The following distribution diagrams illustrate the \textit{Profit} and \textit{ROI} attributes before and after noise reduction (removal of extreme outliers), as shown in Figures~\ref{fig:roi_outliers}, \ref{fig:roi_wo_outliers}, \ref{fig:profit_outliers}, and \ref{fig:profit_wo_outliers}.

Notably, ROI values were capped at approximately 4.99 after outlier removal, reducing the influence of extreme values. This preprocessing step improved the stability and generalization of subsequent analyses.

Building on the cleaned dataset, two data mining techniques were employed: K-Means clustering for pattern discovery and Random Forest classification for predictive analysis. K-Means was selected due to its efficiency in identifying latent structure in numerical datasets and its suitability for partitioning the financial feature space. A limitation of this approach is that it assumes spherical cluster structures and may not capture more complex relationships in the data.

\subsection{Unsupervised Learning: K-Means Clustering}

K-Means clustering was applied to identify latent financial patterns within the dataset. The number of clusters was set to $K = 4$ to enable comparison with the four seasonal categories, although clustering was performed independently of the season attribute. During the segmentation process, the "season" and "tmdb\_id" attributes were excluded to ensure that grouping was based solely on financial characteristics. With "displayStdDevs" and "distanceFunction" enabled, the discovered financial profiles are interpreted as follows:

\begin{itemize}
    \item \textbf{Cluster 0 (Blockbusters):} High-budget profile (approximately \$64M) with an ROI of 1.54.
    \item \textbf{Cluster 1 (Moderate Hits):} Mid-range profile (approximately \$13M budget) with an ROI of 0.99.
    \item \textbf{Cluster 2 (Failures):} Negative profit (approximately -\$7.7M) and ROI (-0.51), representing 51\% of the dataset.
    \item \textbf{Cluster 3 (High Efficiency):} Low-budget profile (approximately \$7M) achieving the highest ROI of 3.32.
\end{itemize}

\begin{figure}[!t]
\centering
\includegraphics[width=\columnwidth]{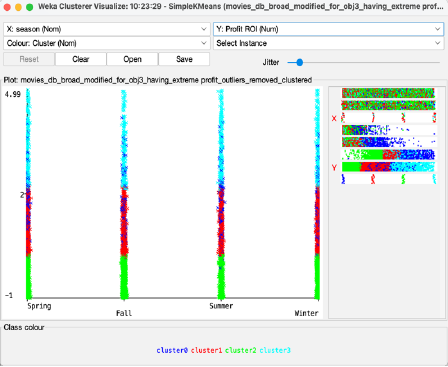}
\caption{Profit Distribution Across Seasons}
\label{fig:profit_season}
\end{figure}

\begin{figure}[!t]
\centering
\includegraphics[width=\columnwidth]{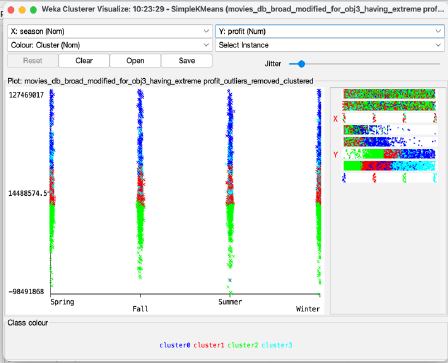}
\caption{ROI Distribution Across Seasons}
\label{fig:roi_season}
\end{figure}

\begin{figure}[!t]
\centering
\includegraphics[width=\columnwidth]{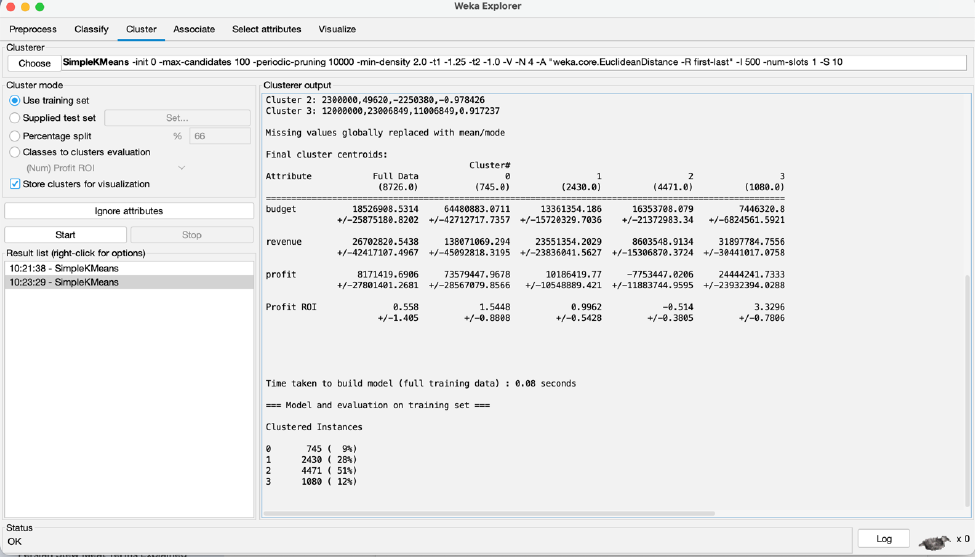}
\caption{K-Means Clustering Output}
\label{fig:kmeans_output}
\end{figure}

The distribution of profit and ROI across seasons is illustrated in Figures~\ref{fig:profit_season} and \ref{fig:roi_season}, while the clustering output is presented in Figure~\ref{fig:kmeans_output}.

Analysis of cluster membership indicates that winter releases are proportionally distributed across these four segments in similar ratios to Spring, Summer, and Fall. This suggests that the winter seasonal category does not exhibit discriminatory power with respect to high-profit segments. Consistent with this observation, the class distribution across seasons is relatively balanced, indicating that differences in predictive performance are not driven by class imbalance.

\subsection{Supervised Learning: Random Forest}

A Random Forest classifier was trained to model the relationship between the financial attribute space and the target class label (season). The tmdb\_id attribute was removed to ensure that the model learned generalizable patterns rather than memorizing individual instances. The experiment used 100 iterations and 10-fold cross-validation.

The classification results indicate that financial attributes provide limited predictive power for determining release season (a = spring, b = fall, c = summer, d = winter). The confusion matrix is presented in Table~\ref{tab:rf_confusion}.

\begin{table}[H]
\centering
\caption{Random Forest Confusion Matrix for Season Classification}
\label{tab:rf_confusion}
\begin{tabular}{lcccc}
\toprule
 & \textbf{a} & \textbf{b} & \textbf{c} & \textbf{d} \\
\midrule
\textbf{a} & 471 & 627 & 410 & 452 \\
\textbf{b} & 570 & 847 & 579 & 614 \\
\textbf{c} & 398 & 624 & 493 & 485 \\
\textbf{d} & 441 & 676 & 468 & 571 \\
\bottomrule
\end{tabular}
\end{table}

Model performance is summarized in Table~\ref{tab:rf_metrics}.

\begin{table}[H]
\centering
\caption{Random Forest Model Performance Metrics}
\label{tab:rf_metrics}
\resizebox{\columnwidth}{!}{
\begin{tabular}{lr}
\toprule
\textbf{Metric} & \textbf{Value} \\
\midrule
Correctly classified instances & 2,382 / 8,726 (27.29\%) \\
Kappa & 0.024 \\
MAE & 0.367 \\
RMSE & 0.4725 \\
\bottomrule
\end{tabular}
}
\end{table}

The evaluation indicates weak predictive performance. The accuracy is 27.29\%, and the Kappa statistic of 0.024 suggests performance only marginally above a random baseline.

The confusion matrix further highlights the absence of a distinct seasonal signature. Winter releases were correctly classified in only 571 instances, with substantial overlap across categories. For example, 676 winter films were misclassified as Fall releases, indicating that films released toward the end of the year share a similar distribution in the financial feature space.

These findings are consistent with the clustering results, which also showed no clear separation of financial profiles across seasons. Together, both analyses suggest that the selected financial attributes are not strongly associated with seasonal release patterns.

Although high-profile films may be strategically released during winter, the dataset reflects a consistent mix of successful and underperforming films across all seasons. This indicates that factors beyond season, such as genre, marketing, or franchise status, are likely more influential in determining movie profitability.

\section{Uncovering Movie Revenue Key Drivers}

This section presents a machine learning and SHAP (SHapley Additive exPlanations) pipeline to uncover patterns and key determinants of film revenue using production and release metadata for 10,895 films. Two regression models, Decision Tree and Random Forest, were trained and evaluated, and SHAP analysis was applied to interpret the key drivers of revenue prediction.

The dataset movies\_db\_broad contains 10,895 films with complete budget and revenue records. Nine predictor variables were selected based on a review of relevant literature, with revenue specified as the target variable. Table~\ref{tab:features_revenue} summarizes the features used in the analysis.

\begin{table}[H]
\centering
\caption{Selected Features for Revenue Prediction}
\label{tab:features_revenue}
\resizebox{\columnwidth}{!}{
\begin{tabular}{ll}
\toprule
\textbf{Feature} & \textbf{Description} \\
\midrule
budget & Production budget of the film \\
rating & Audience or critic rating score \\
popularity & Pre-release popularity score \\
runtime & Length of the film in minutes \\
days\_to\_holiday & Days between release date and nearest holiday \\
days\_oscar & Days between release date and the Oscars ceremony \\
season & Release season (Spring, Summer, Fall, or Winter) \\
nearest\_holiday & The closest holiday to the film's release date \\
genres & Genre classification(s) of the film \\
\bottomrule
\end{tabular}
}
\end{table}

The three categorical variables "season", "nearest\_holiday", and "genres" were converted to numeric values using Label Encoding, as machine learning models require numeric input.

The dataset was partitioned into two subsets to enable unbiased model evaluation:
\begin{itemize}
    \item Training set: 8,716 films (80\%) — used to fit the models
    \item Test set: 2,179 films (20\%) — used to evaluate prediction performance on unseen data
\end{itemize}

A fixed random seed (random\_state = 42) was applied to ensure reproducibility of results.

Model performance was evaluated using two standard metrics:
\begin{itemize}
    \item RMSE (Root Mean Squared Error): measures average prediction error in dollars. Lower values indicate better accuracy.
    \item $R^2$ (R-squared): measures the proportion of revenue variance explained by the model. Values closer to 1.0 indicate better fit.
\end{itemize}

A Decision Tree Regressor was first trained with a maximum depth of 5 to balance model complexity and generalizability. The model produced the results shown in Table~\ref{tab:dt_results}.

\begin{table}[H]
\centering
\caption{Decision Tree Regression Performance}
\label{tab:dt_results}
\resizebox{\columnwidth}{!}{
\begin{tabular}{llp{6cm}}
\toprule
\textbf{Metric} & \textbf{Value} & \textbf{Interpretation} \\
\midrule
RMSE & \$127,218,452 & Average prediction error of approximately \$127M \\
$R^2$ & 0.397 & Model explains approximately 40\% of revenue variance \\
\bottomrule
\end{tabular}
}
\end{table}

These results indicate that the decision tree captures only a limited share of the variation in revenue. The relatively low $R^2$ reflects the model's difficulty in representing the complex and non-linear relationships present in the data.

To improve predictive performance, a Random Forest Regressor was trained using an ensemble of 100 decision trees with a maximum depth of 10. By aggregating predictions across multiple trees, the model reduces variance and improves generalization. The comparative results are presented in Table~\ref{tab:rf_vs_dt}.

\begin{table}[H]
\centering
\caption{Decision Tree vs. Random Forest Performance}
\label{tab:rf_vs_dt}
\resizebox{\columnwidth}{!}{
\begin{tabular}{lcc}
\toprule
\textbf{Metric} & \textbf{Decision Tree} & \textbf{Random Forest} \\
\midrule
RMSE & \$127,218,452 & \$96,582,628 \\
$R^2$ & 0.397 & 0.652 \\
\bottomrule
\end{tabular}
}
\end{table}

The relatively low $R^2$ suggests that the decision tree struggled to capture the complex, non-linear relationships present in the data.

The Random Forest model provides a substantial improvement in predictive accuracy:
\begin{itemize}
    \item $R^2$ increased from 0.397 to 0.652, representing a 64\% relative gain in explanatory power.
    \item RMSE decreased from approximately \$127M to \$96M, a reduction of about \$31M in average prediction error.
\end{itemize}

\begin{figure}[!t]
\centering
\includegraphics[width=\columnwidth]{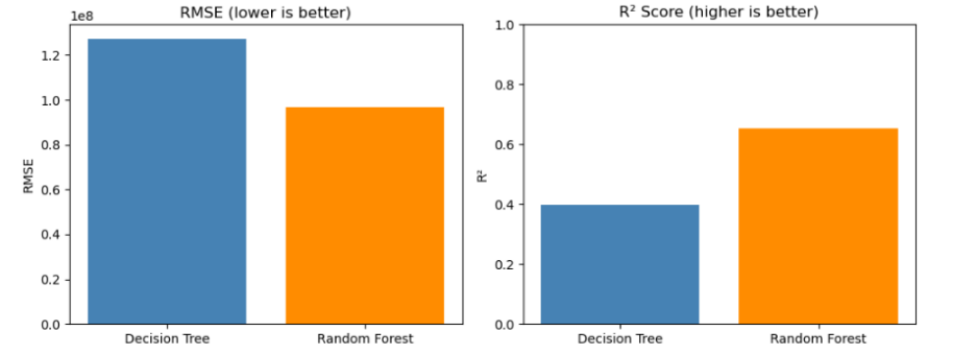}
\caption{Decision Tree and Random Forest Regression Performance Comparison}
\label{fig:decisiontree_rf}
\end{figure}

Figure~\ref{fig:decisiontree_rf} illustrates the performance comparison between the Decision Tree and Random Forest models.

The Random Forest model demonstrates stronger predictive performance than the single Decision Tree. $R^2$ improved from 0.397 to 0.652, representing a 64\% relative gain in explanatory power. RMSE decreased from approximately \$127M to \$96M, reducing the average prediction error by about \$31M.

To further examine the relationship between budget and financial performance, films were classified using median splits for both budget and revenue. The resulting distribution is presented in Table~\ref{tab:budget_revenue}.

\begin{table}[H]
\centering
\caption{Budget vs. Revenue Classification}
\label{tab:budget_revenue}
\resizebox{\columnwidth}{!}{
\begin{tabular}{lcc}
\toprule
\textbf{Budget Group} & \textbf{High Revenue} & \textbf{Low Revenue} \\
\midrule
High Budget & 4,110 & 1,073 \\
Low Budget  & 1,340 & 4,372 \\
\bottomrule
\end{tabular}
}
\end{table}

\begin{figure}[!t]
\centering
\includegraphics[width=\columnwidth]{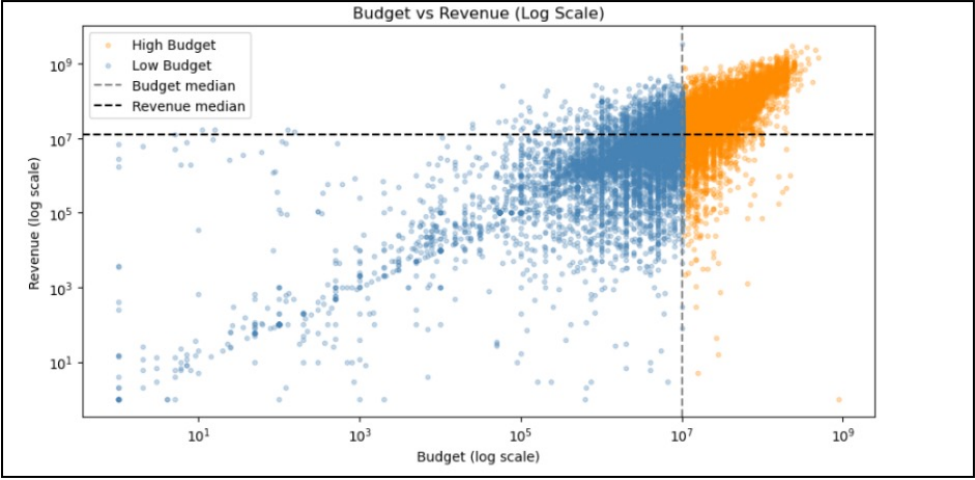}
\caption{Budget vs. Revenue Classification}
\label{fig:budget_revenue}
\end{figure}

A key finding is that 23.5\% of low-budget films achieved above-median revenue. While high budgets remain strongly associated with commercial performance, a meaningful subset of lower-budget productions exceeded revenue expectations. The distribution of films across budget and revenue categories is illustrated in Figure~\ref{fig:budget_revenue}.

This result complements the modeling findings. Both the Random Forest and clustering analyses indicate that financial outcomes are not strongly separated by simple attributes such as season, and instead reflect a distribution across multiple performance profiles. The presence of high-revenue outcomes among low-budget films further reinforces this pattern, suggesting that budget alone does not fully determine financial success.

These findings indicate that additional factors, such as popularity and critical reception, play an important role in shaping revenue outcomes and can offset limited financial investment.

\paragraph{SHAP Feature Importance Analysis}

SHAP (SHapley Additive exPlanations) analysis was applied to the Random Forest model to quantify each feature's contribution to individual predictions. Unlike built-in feature importance measures, SHAP provides directional insight, indicating not only which features matter but also whether they increase or decrease predicted revenue.

\begin{figure}[!t]
\centering
\includegraphics[width=\columnwidth]{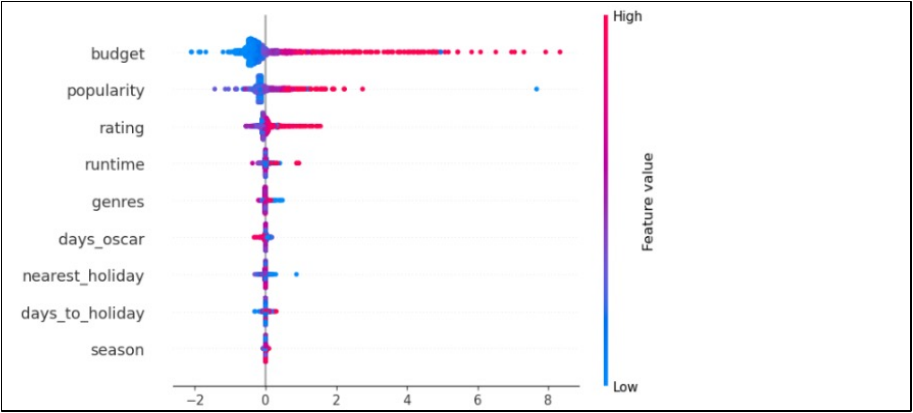}
\caption{SHAP Beeswarm Plot for Revenue Prediction}
\label{fig:shap_values}
\end{figure}

The SHAP beeswarm plot is presented in Figure~\ref{fig:shap_values}.

The beeswarm plot displays one point per film in the test set for each feature. The horizontal position of each point represents its SHAP value, indicating the magnitude of that feature's contribution to the model's revenue prediction relative to the baseline. Points to the right of zero indicate an upward contribution to revenue, while points to the left indicate a downward contribution. The color of each point encodes the original feature value, where red represents high values and blue represents low values.

Reading the plot from top to bottom (most to least impactful), a clear hierarchy of feature importance emerges:

\begin{itemize}
    \item \textbf{Budget:} The widest spread of any feature, with SHAP values ranging from approximately -\$200M to +\$800M. Red points (high budgets) cluster strongly to the right, confirming that films with large production budgets receive substantially higher revenue predictions. Blue points (low budgets) pull predictions downward. The magnitude of this spread clearly indicates that budget is the most influential variable in the model.
    
    \item \textbf{Popularity:} Exhibits a clear red-right, blue-left pattern, although with a narrower spread than budget. High-popularity films consistently receive upward adjustments to predicted revenue, while low-popularity films are associated with downward adjustments. The effect is both directionally consistent and meaningful in magnitude.
    
    \item \textbf{Rating:} Displays a moderate spread, with red points tending toward the right. The relationship is positive but weaker than that of budget or popularity. While higher ratings generally increase predicted revenue, some low-rated films still show positive contributions, indicating variability in this relationship.
    
    \item \textbf{Runtime and Genres:} Both features exhibit narrow distributions centered near zero, indicating limited influence on individual predictions. The color gradients are also less pronounced, suggesting no strong systematic relationship with revenue in this model.
\end{itemize}

Timing-related features ("days\_oscar", "nearest\_holiday", "days\_to\_holiday", and "season") are tightly clustered around zero with minimal spread and no clear color pattern. This indicates that release timing variables have negligible and inconsistent effects on revenue predictions across the test set.

\begin{figure}[!t]
\centering
\includegraphics[width=\columnwidth]{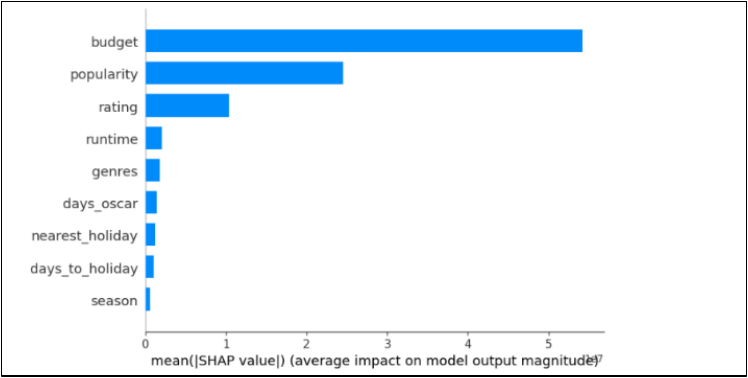}
\caption{Average Absolute SHAP Values by Feature}
\label{fig:avg_shap_values}
\end{figure}

Figure~\ref{fig:avg_shap_values} summarizes the average absolute SHAP values across features, reinforcing the dominance of budget and popularity relative to all other variables.

Budget is the dominant predictor, with an average SHAP impact of approximately \$52M, more than double that of the second most important feature. This confirms that production investment is the strongest driver of box office revenue. Popularity ranks second, with an average impact of approximately \$25M, indicating that pre-release attention and audience awareness play a substantial role in shaping revenue outcomes. Rating ranks third, with an average impact of approximately \$10M, suggesting that critical and audience reception contributes meaningfully to revenue, although its influence is smaller than that of budget and popularity.

In contrast, runtime and genres exhibit minimal impact, each contributing approximately \$2M on average, indicating that film length and genre classification play a limited role in revenue prediction once the primary drivers are accounted for. Timing-related features, including "days\_oscar", "nearest\_holiday", "days\_to\_holiday", and "season", show near-zero impact, indicating that release timing alone is not a meaningful predictor of box office revenue in this dataset.

These findings indicate that a film's financial success is primarily driven by production investment and market attention, rather than content characteristics or release timing.

\section{Conclusion}

This study analyzed the TMDB dataset to identify meaningful associations and relationships among film attributes. The results indicate that film performance is inherently complex and cannot be explained by any single factor. Although production budget is positively associated with financial outcomes, this relationship alone is not sufficient to reliably predict commercial or critical success.

Several commonly held industry assumptions are not supported by the data. In particular, variables such as release timing and runtime show little to no predictive power. This is consistent with prior research suggesting that film performance is shaped by the interaction of multiple factors rather than a single dominant attribute.

These findings have practical implications for decision-making in the film industry. Instead of relying on simple heuristics, producers and studios may benefit from considering the combined effects of budget, marketing, genre, ratings, and popularity. Future research could extend this analysis by incorporating additional variables, such as social media engagement or information on actors and directors, to further improve the understanding of revenue drivers.

This analysis highlights the difficulty of predicting film success, reflecting the multidimensional and uncertain nature of the movie industry.

\newpage

\bibliographystyle{IEEEtran}
\bibliography{references}

@article{daniels1998movie,
  title={Movie money: Understanding Hollywood's (creative) accounting practices},
  author={Daniels, Bill and Leedy, David and Sills, Steven D},
  journal={(No Title)},
  year={1998}
}

@article{seay2025movie,
  title={Movie Sequel Marketing and Predictive Analytics: An Empirical Study on Movie Sequel Marketing Analytics on Marvel’s, Wakanda Forever with Moviegoers},
  author={Seay, EL and Shedrick, Robin and Goodnough, Wanda and Walters, Soun’Ja and Miles, D Anthony and Garcia, Joshua R and Olagundoye, Eniola and Tymann, Nathan and others},
  journal={Anthony and Garcia, Joshua R. and Olagundoye, Eniola and Tymann, Nathan},
  year={2025}
}

@article{drost2021,
  title={The Box Office and the Long Tail: An Examination of the Effects of Streaming on the Distribution of Box Office Revenue},
  author={Drost, Tristan}
}

@article{xiao2024modeling,
  title={Modeling Influencing Factors in US Film Success (1940--2024)},
  author={Xiao, J.},
  journal={Modern Economy},
  volume={15},
  number={12},
  pages={1319--1334},
  year={2024},
  publisher={Scientific Research Publishing}
}

@article{einav2007seasonality,
  title={Seasonality in the US motion picture industry},
  author={Einav, Liran},
  journal={The Rand journal of economics},
  volume={38},
  number={1},
  pages={127--145},
  year={2007},
  publisher={Wiley Online Library}
}

@article{lash2015,
  author       = {Michael T. Lash and
                  Kang Zhao},
  title        = {Early Predictions of Movie Success: the Who, What, and When of Profitability},
  journal      = {CoRR},
  volume       = {abs/1506.05382},
  year         = {2015},
  url          = {http://arxiv.org/abs/1506.05382},
  eprinttype   = {arXiv},
  eprint       = {1506.05382},
  bibsource    = {dblp computer science bibliography, https://dblp.org}
}

\end{document}